# Retrospective clinical evaluation of a decision-support software for adaptive radiotherapy of Head & Neck cancer patients

**Sébastien A. A. Gros[1], Anand P. Santhanam[2], Alec M. Block[1], Bahman Emami[1], Brian H. Lee[1] and Cara Joyce[1]**

[1]Loyola University Chicago, Loyola University Medical Center, Stritch School of Medicine, Department of Radiation Oncology, Cardinal Bernardin Cancer Center, Maywood, IL, 60153 USA
[2]Department of Radiation Oncology, University of California, Los Angeles, CA, 90095 USA

* Correspondence :
Sébastien A. A. Gros
sgros@luc.edu

## Abstract

**Purpose:** To evaluate the clinical need for an automated decision-support software platform for Adaptive Radiation Therapy (ART) of Head-and-Neck cancer (HNC) patients.

**Methods:** We tested RTapp (SegAna), a new ART software platform for deciding when a treatment replan is needed, to investigate a set of 22 HNC patients' data retrospectively. For each fraction, the software estimated key components of ART such as daily dose distribution, and cumulative doses received by targets and OARs from daily 3D imaging in real-time. RTapp also included a prediction algorithm that analyzed dosimetric parameters (DP) trends against user-specified thresholds to proactively trigger adaptive replanning up to 4 fractions ahead. The DPs evaluated for ART were based on treatment planning dose constraints. Warning ($V_{95}$<95%) and adaptation ($V_{95}$<93%) thresholds were set for PTVs, while OAR adaptation dosimetric endpoints of +10% ($DE_{10}$) were set for all $D_{max}$ and $D_{mean}$ DPs. Any threshold violation at end of treatment (EOT) triggered a review of the DP trends to determine the threshold-crossing fraction **Fx** when the violations occurred. The prediction model accuracy was determined as the difference between calculated and predicted DP values with 95% confidence intervals ($CI_{95}$).

**Results:** RTapp was able to address the needs of treatment adaptation. Specifically, we identified 15/22 studies (68%) for violating PTV coverage or parotids $D_{mean}$ at EOT. Nine PTVs had $V_{95}$<95% (mean coverage decrease of -7.7 ±3.3%) including 4 flagged for adaptation at median **Fx**=11.5 (range: 6-18). Fifteen parotids were flagged for exceeding $D_{mean}$ dose constraints with a median increase of +3.18 Gy (range: 0.18-6.31 Gy) at EOT, including 8 with DP>$DE_{10}$. The differences between predicted and calculated PTV $V_{95}$ and parotids $D_{mean}$ was up to 7.6% (mean±$CI_{95}$: -2.9±4.6%) and 5 Gy (mean±$CI_{95}$: 0.2±1.6 Gy), respectively. The most accurate predictions were obtained closest to the threshold-crossing fraction. For parotids, the results showed that **Fx** ranged between fractions 1 to 23, with a lack of specific trend demonstrating that the need for treatment adaptation may be verified for every fraction.

**Conclusion:** Integrated in an ART clinical workflow, RTapp aids in predicting whether specific treatment would require adaptation up to four fractions ahead of time.

# 1    Introduction

The use of intensity modulated radiation therapy (IMRT) and volumetric modulated arc therapy (VMAT) techniques to treat Head and Neck Cancer (HNC) enable the delivery of highly conformal radiotherapy (RT) treatments with complex dose distributions. Initial issues in RT delivery, such as patient set-up and localization errors, were addressed by the International Commission on Radiation Units and Measurements (ICRU) in the 1980s and 1990s with the recommendations for the delineation of Gross Tumor Volume (GTV), Clinical Target Volume (CTV) and Planning Target Volume (PTV) structures[1–3], and minimized by the continuous improvement of imaging modalities for patient set-up verification. The latest Image Guided Radiation Therapy (IGRT) solutions have pushed the limits of reducing PTV-to-CTV margins to only a few millimeters[4–6], generating a greater sparing of organs at risk (OARs). However, these smaller margins leave very little room for errors as an inadequate PTV coverage could lead to treatment failure. The proximity of critical structures to GTVs, the change of volume and the displacement of targets and OARs, and weight loss during treatment, are now the new challenges faced by radiation oncologists, as they all constitute risks for target under-dosage, leading to possible local failure, or radiation toxicities [7–11]. Successful strategies to improve HNC patients' quality of life after RT include sparing the parotid glands (PGs) and the mandible to decrease the risks of xerostomia[12] and osteoradionecrosis[13].

Adaptive radiotherapy (ART) involves all methods that aim to adapt RT treatments and delivered dose distributions to any specific patient anatomical changes[14]. The main potential benefits of ART are to ensure the adequate dosimetric coverage of targets and to limit OARs doses throughout treatment, assuming that it will increase the therapeutic ratio and provide better outcomes for cancer patients[15]. ART encompasses offline, online, and real-time strategies to mitigate the effects of anatomic shifts and set-up errors[16]. While both offline and online ART involve imaging to review the current anatomy and to assess the need for a new plan, most common ART methods allowed by current technologies for HNC patients are variants of offline strategies[17–22], as these are well suited for the slow progressing nature of anatomic changes observed during HNC RT treatments[8,20,23]. Online methods are more appropriate to address the effects of stochastic patient set-up errors[24,25], such as those caused by the interfractional variation in shoulder position[26,27] or the loose fitting of the immobilization mask due to weight loss. They are however only commercially available for HNC treatment on dedicated adaptive RT systems. Real-time ART introduces additional sophisticated patient monitoring to correct intra-fractional anatomic shifts in real time during treatment delivery[28] and could seem excessive for HNC treatments. In contrast, offline ART can readily be implemented with the current basic clinical RT treatment resources.

Clinical offline ART workflows follow four key steps: imaging, assessment, re-planning, and quality assurance (QA)[16]. It is recommended that HNC patients be monitored with CT or CBCT imaging acquired frequently (daily or weekly) throughout treatment[29,30]. These images are reviewed by a radiation oncologist who then decides whether the treatment plan is to be adapted. The re-planning decision is based on an adaptive review strategy which typically consists of the assessment of anatomic variations and of their impact on the dosimetry of tumor targets and OARs. After the registration of the periodic CBCT images to the initial plan CT, an initial qualitative evaluation visually compares the structure contours from the periodic CBCT images against pre-treatment volumes. A subsequent quantitative assessment requires specialized software tools to compute similarity and distance metrics between the initial volumes and the new structures, and to estimate the treatment dose from the most current patient anatomy. The determination of patient- and plan-specific thresholds based on treatment site, fractionation, and outcome is key to optimize the ART workflow, and to provide an individualized approach that is best suited for HNC patients. However, the ART tasks performed after periodic patient imaging requires several hours of expert physicians, physicists, and dosimetrists, trained in their institutional ART workflow and consume resources that most radiation oncology facilities cannot afford. As no commercial automated ART offline workflow is yet available with gantry-mounted linacs, quantitative changes for targets and OAR structures of interest cannot be estimated in a feasible time for the majority of HNC patients. The time-consuming nature of offline ART workflows lead to delays until the new plan is available for treatment. Consequently, clinicians might continue to administer radiotherapy according to the original treatment plan, which may reduce the efficacy of the radiotherapy treatment sought by triggering a plan adaptation. Therefore, there is a current need for an automated and quantitative framework that will process the daily imaging, automatically generate the contoured structures, compute the dose to be received by the structures of interest, and predict if a re-planning is required to maintain the current plan quality. Such automated workflow would ideally include the implementation of predictive models, to allow for the instantiation of clinical adaptive replanning ahead of time.

This manuscript reports on the performance of a newly developed commercial decision-support software platform for ART, RTapp[TM] (SegAna, Orlando, FL), which automatically tracks and analyses daily anatomical changes throughout an entire course of RT and predict when treatment plans will exceed dose constraints. Most software tasks are optimized to run on a Graphics Processing Unit (GPU) and allow the presentation of the results in near real-time[19]. A retrospective analysis of HNC patient treatment data was performed with RTapp to assess if they would have been candidates for treatment adaptation, based on a set of PTV coverage and OAR dose thresholds defined for triggering adaptive re-planning.

## 2 Materials and Methods

### 2.1 Retrospective cohort

A cohort of 22 HNC patients treated between January and December 2019 at our institution were randomly selected for retrospective analysis of potential treatment adaptation with RTapp, under IRB approved protocol (# LU213253). The HNC disease sites are listed in Table 1. The treatment planning images were acquired on a 32 slice SIEMENS SOMATOM CT Open AS scanner (Siemens Healthineers, Erlangen, Germany) with a reconstructed slice thickness of 3 mm and metal artefact reduction (MAR) enabled by default. All HNC patients were immobilized with a Q-fix Fiberplast® Portrait S-frame Head and Shoulder thermoplastic immobilization mask (Qfix, Avondale, PA). The GTV and nodal CTV targets were contoured by the treating physician prior to applying 2-3 mm PTV margins defined as follow: High Risk (HR) PTV, Intermediate Risk (IR) PTV and Low Risk (LR) PTV. A dosimetrist contoured all normal structures and OARs. The plan optimization followed the list of dose constraints required by the treating physician for each individual plan. Table 2 lists all dose constraints for our patients cohort and the dosimetric endpoints (DE) which may trigger a plan review or adaption under an ART workflow.

*Table 1: Distribution of HNC sites from the patient cohort.*

| Diagnoses per Sites | cases |
|---|---|
| Pharynx | 15 |
| Oral Cavity | 2 |
| Larynx | 3 |
| Salivary glands | 1 |
| Lymph nodes | 1 |

*Table 2: Treatment planning dose constraints and structure specific dosimetric endpoints (DE). The DEs serve as thresholds to trigger the plan review for adaption in an adaptive radiotherapy workflow with RTapp.*

| Organ /Volume of interest | Parameter | Planning Goal | Warning DE | Adaptation DE |
|---|---|---|---|---|
| PTVs | $V95$ | > 95% Rx dose | 95% | 93% |
| | Hotspot | < 110% Rx dose | 110% | 110% |
| Spinal Cord | $D_{max}$ | < 45 or 50 Gy | planning goal | 10% |
| Brainstem | $D_{max}$ | < 45 or 50 Gy | planning goal | 10% |
| Oral Cavity | $D_{max}$ | No hotspot | 110% | 110% |
| Spared Parotid[a] | $D_{mean}$ | < 20 Gy | planning goal | 10% |
| contralateral Parotid[b] | $D_{mean}$ | N/A | planning goal | 10% |
| Cervical Esophagus | $D_{max}$ | No hotspot | 110% | 110% |
| Mandible | $D_{1cc}$ | 65 to75 Gy | planning goal | planning goal |
| Cochlea | $D_{mean}$ | < 35 Gy | planning goal | 10% |
| Larynx | $D_{max}$ | No hotspot | 110% | 110% |
| Brachial plexus | $D_{max}$ | 65 or 66 Gy | planning goal | planning goal |

[a] If planning goal achieved, DE relative to 20 Gy. If not and $D_{mean} < 21$ Gy, DE are relative to 21 Gy.
[b] If planned PG $D_{mean} < 26$ Gy then DE relative to 26 Gy.
DE = Dosimetric Endpoint

All patients received External Beam Radiation Therapy (EBRT) with VMAT in a Simultaneous Integrated Boost (SIB) setting in 30-35 fractions without treatment adaptation on a Varian Truebeam linear accelerator (Varian, Palo Alto, CA). Patient set-up verification and alignment were performed before each treatment fraction with kV-CBCT acquired with the Varian on-board imaging (OBI) system. Each CBCT image set was comprised of 93 images with 2 mm separation. Table 3 summarizes the distribution of treatment prescriptions and targets included in this study. All cases were selected to ensure that complete PTV and PG volumes were included in the CBCT field of view. Each patient dataset was anonymized prior to be exported as DICOM RT objects for processing by RTapp.

*Table 3: Treatment plans prescriptions from patient cohort. All patients received their treatment in a Simultaneous Integrated Boost (SIB) setting.*

| SIB Doses (Gy) | Dose per fraction (Gy) | Total # fractions | Targets | # cases |
|---|---|---|---|---|
| 60 / 54 | 2 / 1.8 | 30 | Post-operative bed / nodal basin | 7 |
| 70 / 63 / 56 | 2 / 1.8 / 1.6 | 35 | GTV / CTV / nodal basin | 14 |
| 66 / 60 / 54 | 2.2 / 2 / 1.8 | 30 | GTV / CTV / nodal basin | 1 |
| 66 / 54 | 2.2 / 1.8 | 30 | GTV / CTV | 1 |
| 59.4 / 54.12 | 1.8 / 1.64 | 33 | High risk mucosa & ipsilateral node basin / at risk contralateral nodes | 1 |

## 2.2   Overview of the adaptive software platform

### 2.2.1   RTapp workflow

The main purpose of RTapp is to estimate the dose received by each structure at any treatment time point. A predictive algorithm analyzes the trend of structure specific dosimetric parameters (DP) against pre-determined dosimetric endpoint (DE) values to forecast if, and so when, any dose constraint would be violated. The automated RTapp workflow can be divided in 3 steps, as summarized in Figure 1.

**Step 1.** For each treatment fraction, the initial treatment plan CT images and structures are deformed to match the daily set-up CBCT images. An optical flow based deformable image registration (DIR) algorithm[31] automatically registers the initial and daily 3D image sets and generates a deformation vector field (DVF).

**Step 2.** The DVF is then employed to deform the initial plan structures and dose into daily deformed structures and dose. The deformation results can be verified via DIR confidence metrics.

**Step 3.** The daily dose distribution within any structure is calculated from the deformed structures and dose. The Dose Volume Histograms (DVH) for the day of treatment ($DVH_{day}$) and up to the current treatment time are generated from the estimated dose distribution (Fig. 2G). The day of treatment DVH represents the daily dose scaled to the whole course of treatment, assuming that the daily anatomy would be maintained for the whole course of treatment. The sum DVH ($DVH_{sum}$) summarizes structures doses accumulated up to the current fraction, scaling the latest fraction dose to the remaining course of treatment. It assumes that the most current structure anatomy and dose distribution would hold for the remaining treatment fractions. Specific dosimetric parameters $DP_{day}$ and $DP_{sum}$, are calculated from the $DVH_{day}$ and $DVH_{sum}$ to populate a dose trend graph displayed on the front end of RTapp (fig. 2H). A linear predictive model analyzes the trends of $DP_{sum}$ to forecast their values over the next four fractions and provide quantitative data to guide the decision to trigger adaptive re-planning proactively.

### 2.2.2   Implementation environment

RTapp was tested as a standalone application installed on a Microsoft Windows 10 workstation equipped with a 2.3 GHz intel Core i-9 CPU, 32 GB RAM, and a Nvidia GeForce RTX2070 (8 GB RAM). The processing of a single fraction data set (93 CBCT images and ~30 structures) typically took <1 min.

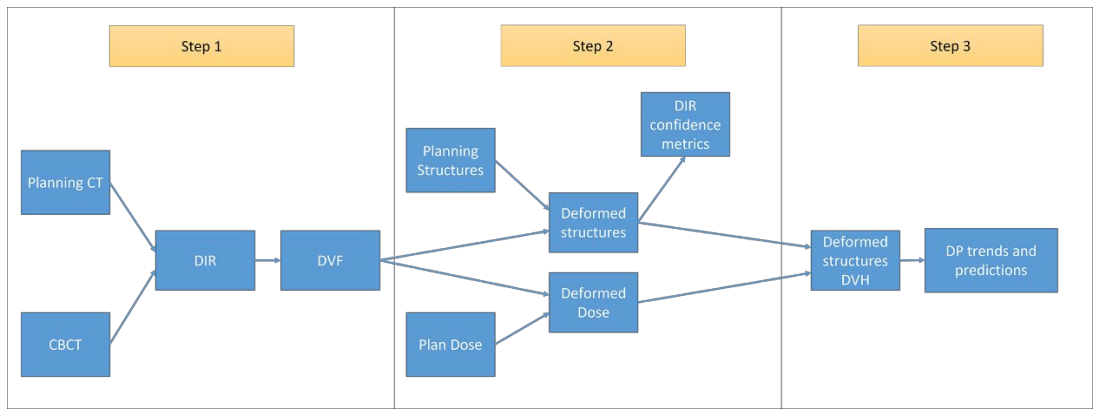

*Figure 1: Schematic of RTapp workflow divided in 3 steps. 1- The deformable registration of planning CT to daily CBCT anatomy generate the Deformation Vector Field (DVF). 2- Deformation of initial planning structures and dose based on daily DVF. 3- Generation of structures DVH based on daily deformed structures and dose with trend and prediction of dosimetric parameters.*

### 2.2.3   Software Front-end

The front-end of the application (Fig. 2) displays treatment specific data, anatomic visualization windows (fig. 2A to F) and panels with graphical data (fig. 2G to I) to guide the ART decision making process for the selected patient study.

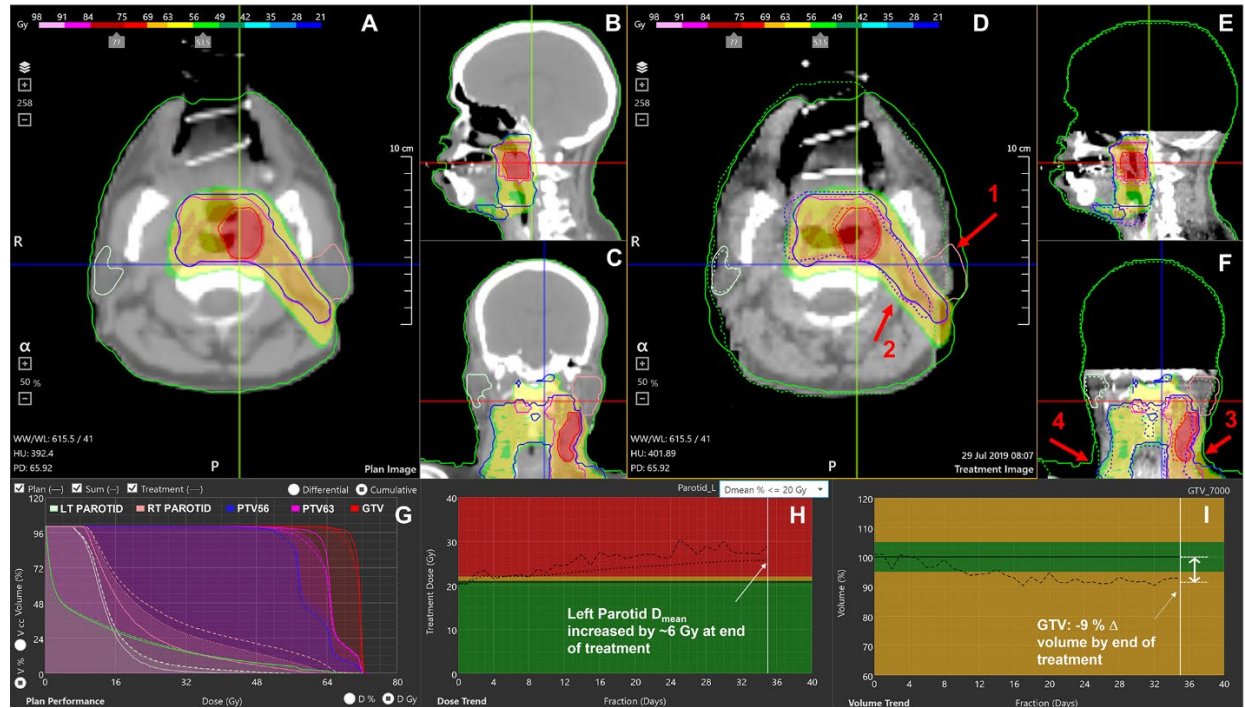

*Figure 2: RTapp software user interface. (A-C) Initial plan CT, structures and 3D dose. (D-F) The final fraction's CBCT with plan structures (solid line), deformed contours to current day's position (dotted line) and initial 3D dose. (G) Dose Volume Histogram showing the plan dose (solid line), delivered dose up to the current fraction ($DVH_{sum}$, dotted line) and the day of treatment dose scaled to full treatment ($DVH_{day}$, dashed line). (H) Trend for the parotid's mean dose and (I) tumor volume regression throughout the treatment. The red arrows indicate notable effects of RT and weight loss. (1) The left parotid regressed and shifted inwards (dotted contour) into the prescribed dose area. (2) The PTVs deformed inwards, thus reducing their dose coverage. (3) The visible effect of weight loss as lateral shrinkage of the green neck contour (dotted line vs. plain line) and the inward shifts of all targets (the blue and purple PTVs) causing their dose coverage to decrease, in addition to inconsistent shoulders repositioning (4).*

### 2.3   Evaluation of DIR quality

The quality of the deformation was first assessed on the CBCT viewing panels (Fig. 2, D-F) by visually comparing overlaid initial and deformed structures contours. A quantitative evaluation was then performed with the two DIR confidence metrics provided by RTapp, following the recommendations from the AAPM TG-132 report[32]. Structures with normalized cross correlation (NCC) values < 0.85 or for which the displacement vector of the secondary image voxels within a structure exceeding a 7 mm 'large' displacement threshold after deformation were automatically flagged for review. The DIR algorithm parameters were adjusted before reprocessing a fraction when the flagged structures deformations were assessed as inaccurate after user review.

### 2.4   Dosimetric metrics and thresholds for Target coverage and OARs

The structures monitored for the retrospective study were all PTV, CTV, GTV and nodal targets, as well as parotid glands (PGs), spinal cord, cochleae, brainstem, mandible, esophagus, and larynx. The DPs evaluated against the need for adaptation were based on the dose constraints required during treatment planning (Table 2). All cases had identical requirements for the PTVs ($V_{95}$ > 95% of prescription dose; maximum dose $D_{max}$ < 110%) and for the PGs ($D_{mean}$ < 20 Gy). Additional PG DPs were defined specifically for this study: $D_{mean}$ < 21 Gy for cases where the dosimetrist could not keep the mean ipsilateral PG dose below 20 Gy due to the overlap with PTVs, and $D_{mean}$ < 26 Gy for spared contralateral PGs. Other OAR constraints varied per plans (Table 2).

## 2.5  Adaptive review strategy

Each patient data set was fully processed with RTapp. The $DVH_{day}$ and $DVH_{sum}$ generated by RTapp for the final dose (Fig 2G) were compared to the initial plan DVH ($DVH_{plan}$).  Any violation of dose constraints at end of treatment (EOT) were tallied as potential case for adaptation. For every dose constraint violation, the $DP_{sum}$ trend (Fig. 2H) was reviewed to determine the fraction when the violation occurred. For this work, a hypothetical "warning" threshold ($V_{95}$ < 95%) was set to investigate the impact of daily set-up variation on PTV coverage and an "adaptation" threshold of -2% ($V_{95}$ < 93%) was set to trigger a review of this patient's anatomy for adaptation. An OAR "adaptation" dosimetric endpoint of 10% ($DE_{10}$) was set uniformly for all $D_{max}$ and $D_{mean}$ DPs.

## 2.6  Prediction model accuracy

The accuracy of the prediction model was evaluated by calculating the difference between the $DP_{sum}$ value at the fraction when it violates the adaptation threshold and the predicted $pDP_{sum}$ values from the four fractions preceding the violation time point. The difference in $DP_{sum}$ was averaged over all patient studies flagged for adaptation to calculate the 95% Confidence Interval as summarized by equation 1. For a particular $DP_{sum}$ value violating a threshold at fraction *Fx*, the difference in $DP_{sum}$ from a predicted $pDP_{sum}$ value based on processed fraction [*Fx-i*] with *i* = [4,3,2,1], averaged over all *n* flagged studies is given by:

$$[\Delta DP_{sum}]_{Fx-i} = \frac{1}{n}\sum_n([pDP_{sum}]_{Fx-i,n} - [DP_{sum}]_{Fx}) \qquad [1]$$

Finally, a potential clinical ART workflow integrating RTapp in the treatment of HNC patients was proposed.

# 3  Results

## 3.1  Adaptive review

The retrospective analysis with RTapp reported 15/22 patient studies which failed to meet at least one dose constraint at EOT. The flagged structures were 8 PTV targets and 15 PGs. Other structures DPs remained below their warning threshold values throughout treatments. Box plots summarizing the differences between the $DVH_{plan}$ and the $DVH_{sum}$ derived PTV $D_{max}$ and $V_{95}$, PG $D_{mean}$ and spinal cord $D_{max}$ are shown in Figure 3. Overall, PTVs dosimetry and coverage decreased while PG and spinal cord doses increased during RT, with the latter remaining below thresholds.

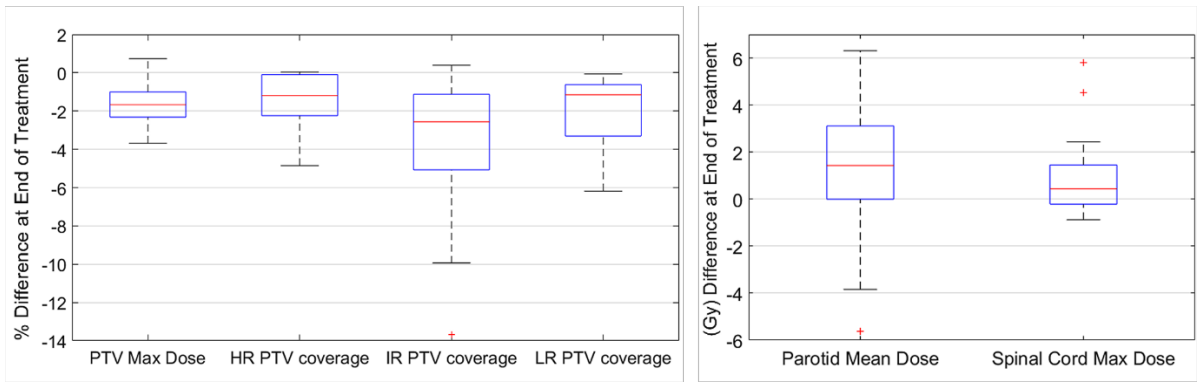

*Figure 3: Boxplot summary of the DP values differences between start and end of treatment for PTVs, PGs and Spinal cord.*

### 3.2 Targets coverage

The difference in $V_{95}$ between planned and EOT $DP_{sum}$ values ranged from 0 to -4.9% for HR PTVs, +0.4 to -13.7 % for IR PTVs and -0.1 to -6.2% for LR PTVs. The difference in PTV $D_{max}$ ranged from +0.74 to -3.7 Gy. While all PTVs met the minimum $V_{95}$ > 95% coverage requirement after treatment planning, eight PTVs belonging to six patients were flagged for under-coverage ($V_{95}$<95%) at EOT. The changes in coverage for these targets are summarized in Table 4. The mean decrease in coverage was $\Delta\overline{V}_{95}$ = (-7.0 ± 3.3) %, resulting in a mean final $\overline{V}_{95}$ = (91.7 ± 3.0) % for the 8 PTVs. Four flagged targets were IR PTVs, accounting for the initial GTVs, involved nodal basin and areas of microscopic spread (studies 94, 118, 146 and 18). Two flagged PTVs were covering post-operative beds (studies 8 and 99) while the last two flagged PTVs were covering low risk nodal basin (studies 8 and 18). The example shown in Fig. 2, from study 118, illustrates the effect of internal anatomical changes and weight loss on the last fraction CBCT (Fig. 2 D-F). The inwards shift and the regression of the deformed PTV contour is responsible for the loss of coverage (dotted line $DVH_{sum}$ on Fig. 2G) for IR PTV63 with an EOT $V_{95}$ value of 92.9%.

*Table 4: PTV coverage for patient studies with $V_{95}$ < 95% coverage at end of treatment. The fraction at which the coverage crossed the 95% warning threshold, and the corresponding cumulative dose were identified from the $V_{95}$ trend.*

| Study # | Target | Initial $V_{95}$ | Final $V_{95}$ | $\Delta V_{95}$ | Fraction for $V_{95}$ < 95% | Cumulative dose (Gy) |
|---|---|---|---|---|---|---|
| 94 | IR PTV60 GTV+nodes | 100.00% | 93.74% | -6.26% | 10 | 20 |
| 118 | IR PTV63 GTV+nodes | 99.43% | 92.86% | -6.57% | 16 | 32 |
| 146 | IR PTV63 GTV+nodes | 99.43% | 93.55% | -5.88% | 1 | 2 |
| 8 | PTV60 Post-op bed | 98.12% | 84.45% | -13.67% | 9 | 18 |
| 8 | PTV54 Nodal basin | 98.36% | 92.17% | -6.19% | 13 | 26 |
| 99 | PTV60 Post-op bed | 94.67% | 93.27% | -1.40% | 1 | 2 |
| 18 | IR PTV63 GTV+nodes | 99.50% | 89.56% | -9.94% | 1 | 2 |
| 18 | LR PTV56 GTV+nodes | 99.97% | 93.81% | -6.16% | 3 | 6 |

## 3.3   Trend analysis of Targets dosimetric parameters

The threshold-crossing fraction for $V_{95} < 95\%$ threshold were extracted from the automated fraction processing reports (Table 4, "Fraction for $V_{95} < 95\%$"). The results hinted at a clustering of PTV coverage constraint violation occurring either during the first three fractions (early), or during the second quarter of the treatment (late), with median final $V_{95}$ values of 93.4% and 92.5%, respectively. However, an Independent Samples 2-tail t-test for equality of the mean final $V_{95}$ values of both groups indicated no significant difference (p = 0.394). The fractions at which a specific PTV crossed the $V_{95} < 93\%$ "adaptation" threshold were determined from their respective $V_{95}$ $DP_{sum}$ trend graphs. Fig. 4A shows an example dose trend from study 118 where the PTV coverage decreases during treatment and crossed the 93% "adaptation" threshold (indicated by the red background) at fraction #26, as pointed out by the arrow. Four PTVs (study 8: PTV60 and PTV54, study 18: PTV56, and study 118: PTV63) were flagged for adaptation before EOT with $V_{95} < 93\%$ at a median fraction of 11.5 (range [6-18]).

## 3.4   Parotids mean dose

The difference between the planned PG $D_{mean}$ and the $DP_{sum}$ value estimated at EOT by RTapp ranged from -5.6 to +6.3 Gy, with 77% of the patients showing an increase in overall PG dose. RTapp reported 11 patients (15 flagged PGs) with at least one PG exceeding their dose constraint. Ten studies with initial PG $D_{mean} < 20$ Gy had at least one PG with $D_{mean} > 20$ Gy at EOT. Three cases with an initial PG $D_{mean}$ between 20-21 Gy were reported for exceeding a 21 Gy $D_{mean}$ by EOT. Three additional contralateral PGs for which the initial plan kept the PG $D_{mean} < 26$ Gy were also flagged. The median increase in PG mean dose at end of treatment was +3.18 Gy (range: 0.99-6.31 Gy) for all 15 flagged PGs. The mean difference between start and EOT for PGs violating the 20 Gy, 21 Gy and 26 Gy $D_{mean}$ constraint were +3.30 Gy, +3.49 Gy and +2.51 Gy, respectively.

*Table 5: Parotid glands $D_{mean}$ violations of the warning dose thresholds ($DE_0$) and of the 10% dosimetric endpoints ($DE_{10}$) for triggering adaptation. The difference $\Delta$ $D_{mean}$ is calculated as Final − Initial. The fractions at which $DE_0$ and $DE_{10}$ are reached, and their respective cumulative treatment dose are indicated in the rightmost column. Studies 105, 78, 116, 146, 79 have PGs which did not reach their respective $DE_{10}$ by end of treatment.*

| study # | Initial plan $D_{mean}$ (Gy) | Final RTapp $D_{mean}$ (Gy) | $\Delta$ $D_{mean}$ (Gy) | Laterality | $DE_0$ (Gy) | Fraction for $D_{mean} > DE_0$ | Cumulative dose for $D_{mean} > DE_0$ | Fraction for $D_{mean} > DE_{10}$ | Cumulative dose for $D_{mean} > DE_{10}$ |
|---|---|---|---|---|---|---|---|---|---|
| 46 | 19.66 | 22.25 | 2.60 | Left | 20 | 2 | 4 | 22 | 44 |
| 81 | 19.54 | 25.85 | 6.31 | Left | 20 | 1 | 2 | 1 | 2 |
| 79 | 19.98 | 23.17 | 3.19 | Left | 20 | 1 | 2 | 1 | 2 |
| 137 | 19.63 | 24.00 | 4.37 | Left | 20 | 2 | 4 | 4 | 8 |
| 94 | 19.40 | 20.95 | 1.55 | Right | 20 | 12 | 24 | 20 | 40 |
| 105 | 19.24 | 20.23 | 0.99 | Right | 20 | 21 | 42 | - | - |
| 78 | 19.57 | 21.86 | 2.29 | Left | 20 | 9 | 18 | - | - |
| 116 | 17.62 | 20.29 | 2.68 | Right | 20 | 16 | 32 | - | - |
| 83 | 19.74 | 25.44 | 5.7 | Left | 20 | 2 | 4 | 2 | 4 |
| 118 | 20.76 | 25.72 | 4.97 | Left | 21 | 2 | 4 | 9 | 18 |
| 137 | 20.21 | 23.77 | 3.56 | Right | 21 | 2 | 4 | 5 | 10 |
| 146 | 20.42 | 22.37 | 1.95 | Right | 21 | 1 | 2 | - | - |
| 81 | 25.66 | 29.14 | 3.48 | Right | 26 | 5 | 10 | 23 | 46 |

| 79 | 25.24 | 26.93 | 1.69 | Right | 26 | 12 | 24 | - | - |
| 105 | 25.47 | 27.82 | 2.35 | Left | 26 | 1 | 2 | - | - |

### 3.5  Trend analysis of Parotids dosimetric parameters

Table 5 summarizes all dose constraints ($DE_0$) and $DE_{10}$ violations for the parotids. The fraction at which the PGs mean doses exceeded their respective 10% deviation thresholds were obtained from the automated fraction processing reports (Table 5, "fraction for $D_{mean} > DE$"). The flagged studies can be divided into 2 independent groups. (1) Eight patient studies had a PG $DP_{day}$ failure occurring within the first two treatment fractions, with an average PG mean dose difference $\Delta D_{mean}$ = 4.01±1.44 Gy. These occurred too early to result from radiation treatment and were most likely due to patients relaxing in their immobilization mask, leading to inconsistent patient set-up throughout the course of treatment. (2) The second group comprises seven studies with endpoints failures occurring later during treatment (median threshold-crossing fraction Fx=20, range: 8 – 22), with an average PG mean dose difference $\Delta D_{mean}$ = 1.84±0.66 Gy, likely to result from gradual weight loss and internal anatomical shifts induced by radiation treatment. An Independent Samples 2-tail t-test for equality of the mean PG dose difference between the early and late groups was significant (p = 0.002). Nine PGs were flagged for adaptation with PG $D_{mean} > DE_{10}$ before EOT, with average PG mean dose differences $\Delta D_{mean}$ = 5.05±1.26 Gy and $\Delta D_{mean}$ = 3.14±1.45 Gy for the early (N=5) and late (N=4) groups, respectively.

### 3.6  Predictive model

Fig. 4C and 4D show the comparison of the model prediction from study 81 (right PG), where the 26 Gy $DE_0$ was exceeded at fraction 8 (Fig. 4B). The predicted trend at fraction 5 (extended dotted line to the right of the white vertical line on Fig. 4C) indicated that the right PG $D_{mean}$ would cross 26 Gy at fraction *Fx*=8. The accuracy of the prediction model was estimated for all flagged studies except those with identified Fx < 5 (Table 4 and 5) as the model requires 5 treated fractions to generate predictions. The overall difference between measured and predicted PTV $V_{95}$ ranged from -7.6 to 0.0% (mean±$CI_{95}$: -2.9±4.6 %), with the largest differences observed for predictions made three fractions ahead (mean: -3.0%, $CI_{95}$: -8.4%, 2.4%) and the most accurate predictions (mean: -2.4%, $CI_{95}$: 0.9%, -5.8 %) obtained closest to the threshold-crossing fraction. Figure 5A summarizes the mean $V_{95}$ difference results as a function of temporal proximity defined as the time interval between the threshold-crossing fraction (*Fx*) and the last processed fractions (*Fx-i*) providing the prediction $V_{95}$[*Fx-i*] with *i* progressing from 4 to 1. Uncertainties are reported as 95% confidence intervals ($CI_{95}$). All model predictions overestimated $V_{95}$ coverage values. The variation in prediction accuracy for the PG $D_{mean}$ is presented in Figure 5.B. The overall difference between the measured and predicted PG $D_{mean}$ values ranged from 0.0 to +5.0 Gy (mean±$CI_{95}$: 0.2±1.6 Gy). The largest differences were calculated for predictions four fractions ahead (mean: 0.65 Gy, $CI_{95}$: -1.83 Gy, 3.13 Gy), while the highest accuracy was obtained within two to one fraction ahead (mean: 0.14 Gy, $CI_{95}$: -0.93 Gy, 1.22 Gy).

### 3.7  Performance of DIR algorithm

The automatically calculated NCC and distance confidence metrics quantitatively identified the structures with questionable deformations and helped to confirm our observations from the qualitative visual review. After several iterations, a single DIR algorithm configuration was found to optimally to process all patients data sets without user adjustment. Fourteen patients from our cohort had dental implants which created streak and beam hardening artifacts on daily CBCT images. The registrations seemed robust against imaging artifacts for all 14 patient's PTV and mandible contours deformations.

Figure 6 illustrates the deformation accuracy of a PTV, mandible and parotid contours in the presence of severe artifacts. The only issues encountered when reviewing DIR results were the misidentifications of the immobilization mask for the external body contour. These mainly occurred during the second half of treatments, after patients' weight loss created air gaps between the skin surface and the shell of the mask.

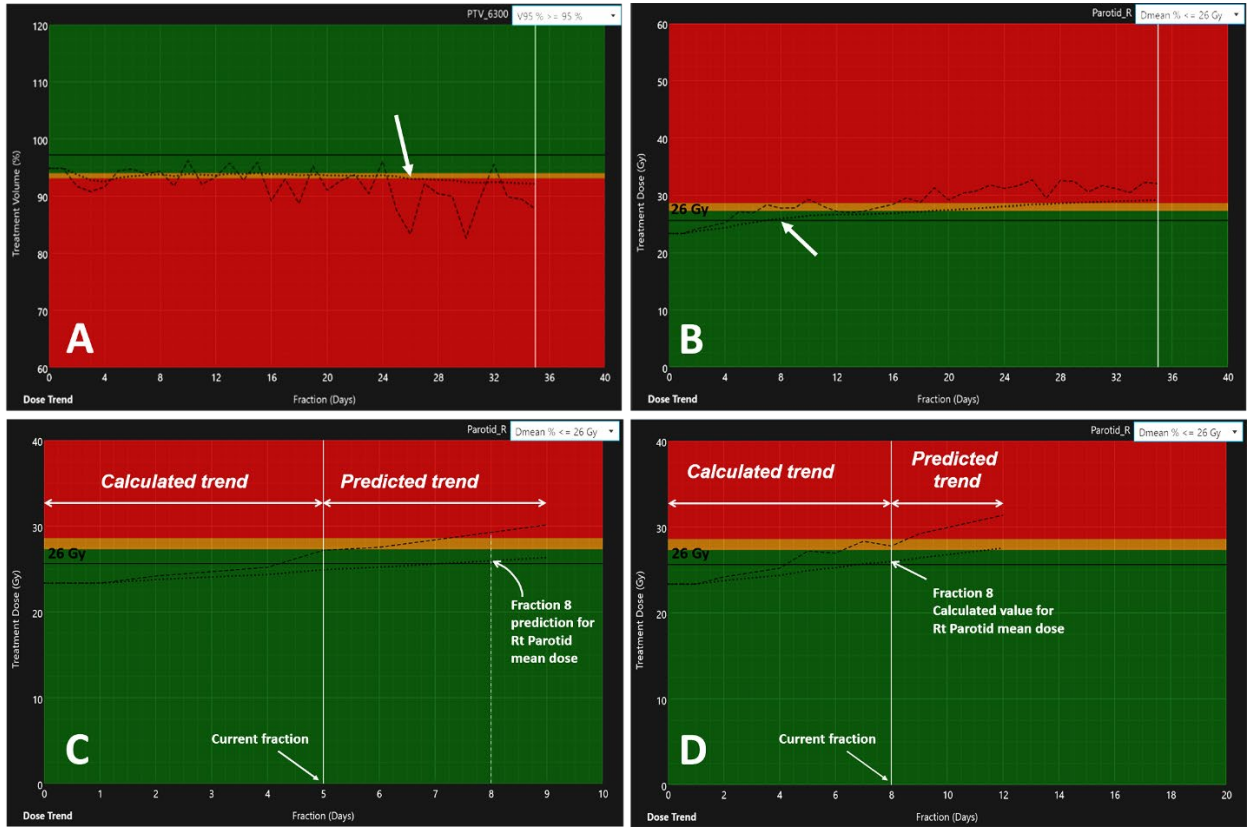

*Figure 4: variation of dosimetric parameters over the full treatments for (A) the $V_{95}$ trend of PTV63 (study 118), and (B) the increase in PG $D_{mean}$ for the Right PG (study 81). The dashed lines indicate the $DP_{day}$ value estimated from the day of treatment $DVH_{day}$, while the dotted line represents the variation of the $DP_{sum}$ estimated from the cumulative $DVH_{sum}$. The arrows indicate the fraction at which the $DP_{sum}$ values reached their respective adaptation thresholds: $V_{95} < 93\%$ at fraction 26 for PTV63 (A) and PG $D_{mean} > 26$ Gy at fraction 8 (B). Panels C and D illustrate the prediction of the trend in $DP_{sum}$ for the Right PG. The data to the left of the current fraction represent calculated values based on daily deformed anatomy. The data to the right predict the variation of the $DP_{day}$ and $DP_{sum}$ values of the PG $D_{mean}$ for the next 4 fractions. In this example (C) the model predicted that the PG $D_{mean}$ would cross the 26 Gy DE at fraction 8 based on the data from fractions 1-5. (D) The PG $D_{mean}$ value calculated from the daily anatomy at fraction 8 is within 0.5 Gy of that predicted at fraction 5.*

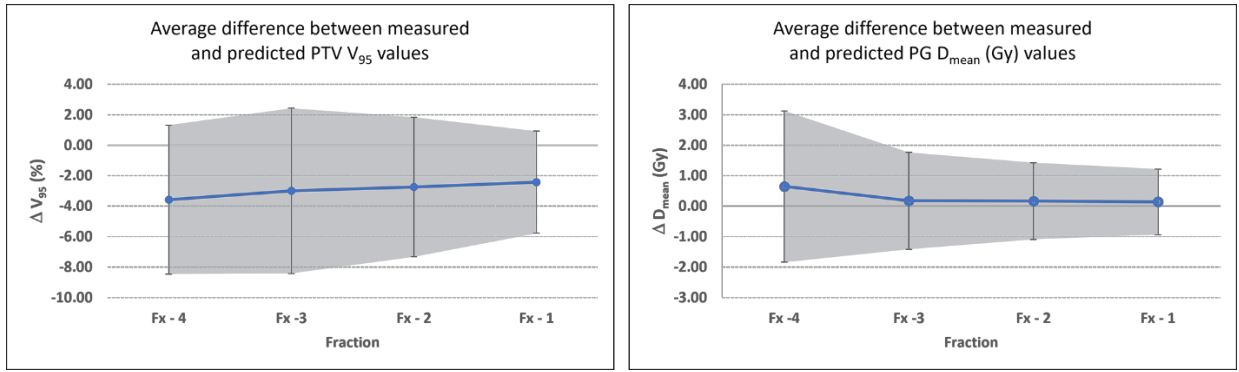

*Figure 5: Differences between measured and predicted values at threshold-crossing for flagged studies for (A) PTV $V_{95}$ and (B) PG $D_{mean}$. Fx is the identified threshold-crossing fraction. FX-i represents the last fraction that was processed to obtain the model predicted value and progresses from Fx-4 to Fx-1 since the model only predicts any $DP_{sum}$ value up to 4 fractions ahead. In this case, Fx-4 represents the first fraction for which a predicted $pDP_{sum}$ value was available while Fx-1 identifies the fraction that directly precedes the threshold-crossing fraction. The confidence bands and error bars indicate the 95% confidence interval.*

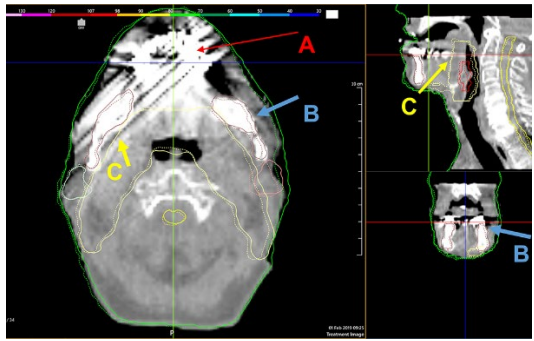

*Figure 6: A demonstration of the deformable algorithm's ability to account for severe CBCT dental implant artifacts. The red arrows (A) point to dental implants artifacts. The blue (B) arrow demonstrates the ability of the DIR to ignore metal scatter for the mandible, a structure with high contrast relative to the surrounding soft tissue. The yellow arrows (C) demonstrate the robustness of the current DIR algorithm, as the auto-segmentation of the yellow PTV contour is unaffected by the streak artifacts.*

## 4   Discussion

This manuscript is the first to report on an automated platform-agnostic commercial software (RTapp) designed to provide quantitative data to support the adaptive re-planning decision process. The first version of RTapp was tested with the retrospective analysis of 22 HNC patients and flagged 15 patient data sets (68%) for failing to meet the minimum required $V_{95}$ PTV coverage or exceeding a PG $D_{mean}$ constraint during treatment. The review of registrations between the daily set-up CBCT and the initial planning CT revealed clear evidence of gradual weight loss and internal anatomical changes during treatment for all flagged patients. Our observations agree with past retrospective and prospective studies of ART for HNC which reported on the reduction of target coverage[9,25,26], and on the increase in PG dose[18,33] and spinal cord dose[9,26] without adaptation. It is highly likely that these 15 patients would have dosimetrically benefited from ART if information on the dosimetric impact of anatomical changes was available during treatment.

The temporal variation of the user defined $DP_{day}$ and $DP_{sum}$ displayed in the front-end of the application can support the adaptive decision process in two ways: (1) Patients for which any tracked $DP_{sum}$ fails to meet a pre-defined adaptation threshold for one or several consecutive fractions might benefit from a re-plan. Results from our patient cohort seem to indicate that HNC patients may be classified into early and late thresholds violation groups. Such dichotomy was statistically significant for PG violating their respective planning constraints ($DE_0$ in table 5). Since the early group violations occurred too early to result from RT, they might be due to patients having issues with immobilization during CT simulation who then relaxed in their immobilization mask during the first few fractions, leading to inconsistent set-up throughout the course treatment. The late group likely includes patients with gradual weight loss and internal anatomical shifts induced by radiation treatment. This classification was statistically inconclusive for PTV $V_{95}$ warning threshold and PG adaptation threshold violations, mainly due to the low sample sizes. (2) A large variability observed in the $DP_{day}$ trend might indicate set-up reproducibility issues. In Fig. 4A, the increasing $DP_{day}$ variability observed after fraction 14 illustrates the effect of gradual patient weight loss on patient set-up and $V_{95}$ coverage. A review of the pre-treatment set-up images revealed large daily variations in head rotation and shoulder positions, indicating that the patient was able to gradually move within her immobilization mask, leading to increasingly set-up reproducibility. Re-planning this patient with a better fitting mask would have improve set-up reproducibility, in addition to raising the PTV coverage and decreasing the $D_{mean}$ to the left PG.

## 4.1   Predictive model

The prediction model implemented in RTapp currently relies on a linear regression to forecast the temporal changes of all $DP_{sum}$ values. Our estimation of the model accuracy for predicting $V_{95}$ coverage values suffers from the low number of flagged studies and resulted in large $CI_{95}$, but still suggested that the model would most likely overestimate $V_{95}$ values. Clinically, this would translate into triggering additional review of target dosimetry in an ART workflow when decisions to adapt are made proactively. The accuracy of the model for the PG $D_{mean}$ with a largest $CI_{95}$[*Fx-4*] of -1.83 to 3.13 Gy would allow the definition of a $2 - 3$ Gy deviation threshold for plan review based on an initial planned PG $D_{mean}$ in the 20 $- 26$ Gy range. Recent decision support methods developed to identify HNC patients that might benefit from adaptation necessitated large patient cohorts to build prediction models for anatomical changes[34,35], tumor response[34,36] and OARs dose accumulation [35,37]. The approach introduced by McCulloch and collaborators[37] was the closest to that implemented in RTapp. Their model predicts the value of specific dose metrics at EOT based on accumulated dose re-calculated from daily CBCT anatomy. Their model achieved > 95% sensitivity and specificity to detect a need for adaptation when predictions were based on a minimum of 10 and 15 treated fractions. Their method involved time-consuming manual steps to reach the prediction result and only provides the dose deviation between planned and received dose at a single time point. In contrast, RTapp generates a prediction for every fraction in real-time, without user intervention, and only relies on the current patient treatment data to build its prediction model. The simple linear regression approach works well for smooth and gradual changes, however the nonlinear treatment response by targets and OARs, in addition to the potentially large variance observed in $DP_{day}$ values, can lead to large errors in the model estimations. Implementing a multiple regression model, which incorporates additional parameters related to patient volume changes, individual structures volume changes or displacements will likely improve the accuracy of the predictions and reduce the error in the model estimation. Ultimately, the implementation of a machine learning based prediction method might provide the most accurate trend of OAR and Targets DPs.

## 4.2   Selection of appropriate dose metrics and deviation thresholds

The "warning" and "adaptation" threshold for specific DPs are at the core of the automated process to initiate a physician review and trigger adaptive re-planning. The hypothetical limits used in this work were based on our physicians constraints provided for treatment planning. While these were helpful to investigate RTapp's approach to measure the dosimetric effects of internal anatomical changes, there is no consensus on which DPs and deviation thresholds are the most appropriate for triggering adaptation. McCoullouch et al[37] used a PG $D_{mean}$ of 24 Gy with a limiting threshold of 15 % (3.6 Gy) based on published NTCP curves[38] and results from a prospective study correlating saliva output with dose received by the PGs[39]. Lee et al[35] analyzed the detection accuracy of their prediction model using the planned PG $D_{mean}$ with a limit deviation threshold of 10% based on the results from Wu et al[20], but also included 7.5% and 5% under the justification that "physicians would welcome any additional sparing of the PGs". Brouwer et al[40] presented a pre-treatment method to select patients for adaptation, based on a PG $D_{mean}$ threshold of 22.2 Gy and deviation of 3 Gy with a near 80% sensitivity. Other approaches could include the 95% ICRU[3] dose-volume recommendation for minimum PTV coverage or use TCP and NTCP derived threshold values for tumors and OARs with specific endpoints. Regardless of the adaptation threshold selected, RTapp supports the monitoring of multiple dose or volume metrics against any threshold on a per-fraction basis, which ultimately allows the treatment adaptation decision to be quantitatively based on actual patient-specific and daily monitored parameters.

## 4.3   Flexibility and robustness of the DIR algorithm

The DIR algorithm implemented within RTapp is based on the initial work presented by Qi et al[19], and provides the flexibility to adjust the deformation settings to match individual patient's anatomy. Several iterations were needed to find a set of parameters which ultimately worked for all 22 patients in our cohort. Dental implants artifacts typically hinder the auto segmentation of head and neck structures adjacent to the mandibular and maxillary region. The MAR employed during the reconstruction of the initial planning CT might have had a significant impact in limiting the effect of metal and beam hardening artifacts on the DIR results for our cohort of patients. Future work will address the issues related to the separation of the mask from the external contour.

## 4.4   ART clinical workflow with RTapp

Integrating RTapp in a clinical workflow would provide a fast and automated solution for monitoring patients who could potentially become candidate for adaptive re-planning. Taking advantage of the real-time processing of IGRT data, the clinical workflow and adaptive review strategy can be divided into 3 successive steps prior to treatment delivery, as described in Fig. 7. (1) The accuracy of the deformation is evaluated by DIR metrics, with the option to adjust the DIR parameters and reprocess daily set-up images. (2) The impact of the combined daily set-up errors and anatomical changes on specific structures is reviewed directly at the control console by the treating therapy staff, who can compare $DP_{day}$ values to warning and action thresholds and decide whether to adjust a daily patient set-up. Such information is currently unavailable on gantry mounted Linacs. It is only available on treatment platforms dedicated to ART which can perform dose recalculation within times typically ranging from 15 to 60 min [41,42]. (3) The treatment team can receive an alert to review the trend and predictions of any $DP_{sum}$ if a warning or adaptation threshold is reached. The information from the trend can help the treating physician with the decision to trigger adaptive re-planning or continue to treat with the current plan. Patients' alignment could potentially be improved daily, or treatments could be postponed if adaptive re-planning is deemed urgent.

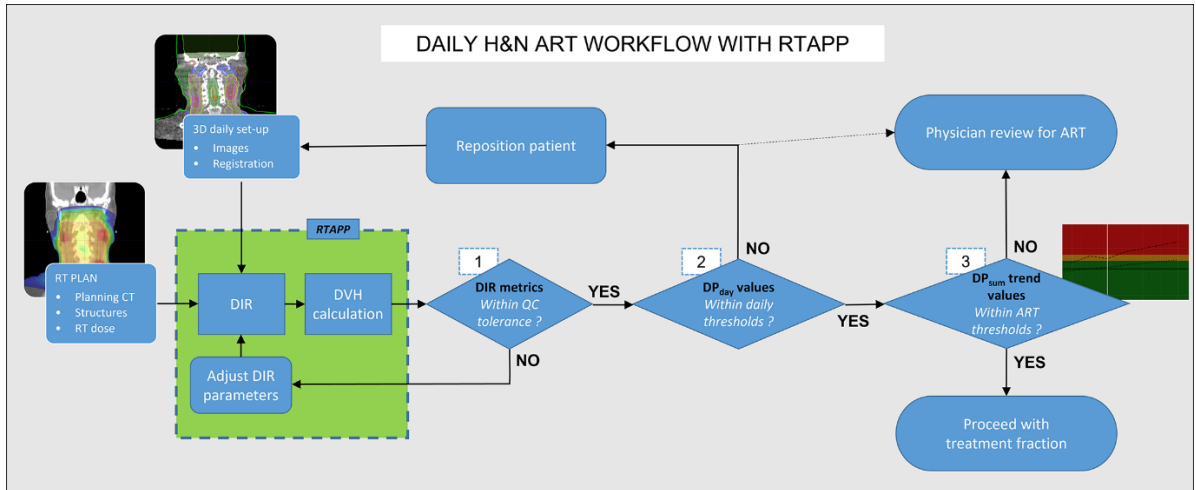

*Figure 7: Hybrid ART clinical workflow with RTapp. The software platform loads daily set-up images as soon as they are available and automatically deforms in the initial plan CT and structures to the daily anatomy. Once the deformation step is completed, RT app calculates DIR metrics, the dose distribution within each structure and tracks DPs. Decision steps 1, 2 and 3 occur at the treatment console prior to treat the patient. (1): RTapp flags structures with potential deformation issues with DIR metrics. (2): The treatment team at the console can review the $DP_{day}$ values against warning and action thresholds and might reposition the patient. (3): the current and predicted trend of $DP_{sum}$ are reviewed prior treatment to assess whether the patient might need treatment adaptation within the next 4 fractions.*

## 4.5   Clinical impact

Online ART is not currently available commercially for gantry mounted linacs. In a typical off-line ART workflow, the successive steps of processing new 3D patient images (including DIR between initial planning CT and most recent CT images), structures re-contouring, re-planning, evaluation of dosimetric and volumetric changes and QA usually take 2 - 5 days. RTapp offers an attractive hybrid solution in between off-line and online ART strategies. It streamlines the off-line ART steps through fast-automated background processes (Fig. 1) and reports dosimetric and volumetric changes in less than a minute, saving hours of dosimetrist, physicist and physician work and optimizing clinical resources in an ART workflow. The implementation of the predictive model would further optimize the use of resources for adaptive re-planning, by allowing to generate a new treatment plan before a patient meets the requirements to trigger adaptation. Compared to recently released fully dedicated on-line ART systems, vendor-agnostic ART decision support platforms provide several advantages. They are readily deployable with any treatment platform equipped with 3D imaging capabilities and make use of resources already available clinically. In addition, their cost-effectiveness is particularly attractive to bring ART to patients from remote rural regions hours away from large academic centers, and from low- and middle-income countries.

The information provided by RTapp to support decision making for ART could be expanded by calculating additional geometric parameters from the daily volumetric data of each deformed structures. Lee et al[43], Yang et al[44] and Surucu et al[33] reported that the tumor volume reduction rate (TVRR) was a prognostic indicator for loco-regional control for oropharyngeal cancers. The work from Barker[8] and Surucu[33] hinted at a strong correlation between weight loss, external contour reduction and volumetric and positional changes in GTVs and normal tissue structures. Gros[45] demonstrated a high correlation between external contour reduction and increase in OARs dosimetry for HNC patients. More recently, Rosen [46] reported that PG volume change and rate of volume change correlated with xerostomia and dosimetric endpoints.

## 5   Conclusion

A novel automated decision support software platform for ART was tested retrospectively with 22 HNC patients' data. Fifteen patients were flagged for adaptation at end of treatment. The trend of PTV coverage and parotid mean doses against specific dose metrics and deviation thresholds on a per-fraction basis demonstrated that RTapp could help identify when to trigger plan adaptation and potentially pro-actively predict when a physician might consider the need for treatment plan adaptation. The tools offered by RTapp have the potential to benefit any clinic equipped with a daily 3D imaging capability without adequate resources to provide ART for their HNC patients. The software platform evaluated provides all the tools and information necessary to design prospective studies aiming to test whether ART will improve outcome both for TCP and NTCP in a diverse range of cancer sites and fractionations.

## 6   Acknowledgements

The authors would like to thank Mr. Saty Seshan (SegAna) for his initial assistance with the RTapp software and Prof. John Roeske (Loyola University Chicago) for comments on the manuscript.